\documentclass[conference,10pt,letterpaper]{IEEEtran}
\IEEEoverridecommandlockouts
\usepackage{graphicx,times,cite,amsmath, amssymb, amsthm, mathrsfs, epsfig, array, subfigure}
\usepackage{multirow}

\usepackage{cite,geometry}

\usepackage{color}

\allowdisplaybreaks[4]
\newcommand{\figwidth}{\linewidth}

\newcommand{\fref}[1]{Fig.~\ref{#1}}

\usepackage[acronym]{glossaries} 

\newacronym{OTFS}{OTFS}{orthogonal time frequency space}
\newacronym{DD}{DD}{delay-Doppler}
\newacronym{TF}{TF}{time-frequency}
\newacronym{AWGN}{AWGN}{additive white Gaussian noise}
\newacronym{PSD}{PSD}{power spectral density}
\newacronym{2D}{2D}{two-dimensional}
\newacronym{BER}{BER}{bit error rate}
\newacronym{ISFFT}{ISFFT}{inverse symplectic finite Fourier transform}
\newacronym{SFFT}{SFFT}{symplectic finite Fourier transform}
\newacronym{FFT}{FFT}{fast Fourier transform}
\newacronym{PDF}{PDF}{probability density function}
\newacronym{MRC}{MRC}{maximum-ratio combining}
\newacronym{SNR}{SNR}{signal-to-noise ratio}
\newacronym{5G}{5G}{fifth-generation}
\newacronym{6G}{6G}{sixth-generation}
\newacronym{UAVs}{UAVs}{unmanned aerial vehicles}
\newacronym{LEO}{LEO}{low earth orbit}
\newacronym{HSRs}{HSRs}{high-speed railways}
\newacronym{OFDM}{OFDM}{orthogonal frequency division multiplexing}
\newacronym{MP}{MP}{message passing}
\newacronym{MAP}{MAP}{maximum \emph{a posteriori} probability}
\newacronym{LMMSE}{LMMSE}{linear minimum mean squared error}
\newacronym{CP}{CP}{cyclic prefix}
\newacronym{AI}{AI}{artificial intelligence}
\newacronym{i.i.d.}{i.i.d.}{independent and identically distributed}

\begin{document}
\title{\LARGE Outage Probability Analysis for OTFS in Lossy Communications}
\markboth{}{}
\author{
\IEEEauthorblockN{Xin Zhang\IEEEauthorrefmark{1}, Wensheng Lin\IEEEauthorrefmark{1}, Lixin Li\IEEEauthorrefmark{1}, Fucheng Yang\IEEEauthorrefmark{2}, Zhu Han\IEEEauthorrefmark{3} and Tad Matsumoto\IEEEauthorrefmark{4}}
\IEEEauthorblockA{\IEEEauthorrefmark{1} School of Electronics and Information, Northwestern Polytechnical University, Xi'an, China \\
\IEEEauthorrefmark{2} Research Institute of Imformation Fusion, Naval Aviation University, Yantai, China \\
\IEEEauthorrefmark{3} University of Houston, Houston, USA \\
\IEEEauthorrefmark{4}IMT-Atlantic, France, and University of Oulu, Finland, Emeritus.  \\
Email: zxx\_@mail.nwpu.edu.cn; \{linwest, lilixin\}@nwpu.edu.cn; fucheng85@sina.com; \\ zhan2@uh.edu; tadeshi.matsumoto@oulu.fi\vspace{-2em}} 
\thanks{This paper has been accepted for publication in IEEE Globecom 2024 workshop.
	
	This work was supported in part by National Natural Science Foundation of China under Grants 62001387 and 62101450, in part by Young Elite Scientists Sponsorship Program by China Association for Science and Technology under Grant 2022QNRC001, in part by Aeronautical Science Foundation of China under Grants 2022Z021053001 and 2023Z071053007, in part by Shanghai Academy of Spaceflight Technology under Grant SAST2022-052,
	in part by French National Research Agency Future Program under reference ANR-10-LABX-07-01, 
	in part by NSF CNS-2107216, CNS-2128368, CMMI-2222810, ECCS-2302469, US Department of Transportation, Toyota and Amazon.
}
}

\maketitle
\begin{abstract}
This paper analyzes the outage probability of orthogonal time frequency space (OTFS) modulation under a lossy communication scenario. First of all, we introduce the channel model and the vector form representation of OTFS this paper uses. 
Then, we derive an exact expression of the OTFS outage probability in lossy communication  scenarios, using Shannon’s lossy source-channel separation theorem.
Because the channel is time-varying, calculating the exact outage probability is computationally expensive. 
Therefore, this paper aims to derive a lower bound of the outage probability, which can relatively easily be calculated. 
Thus, given the distortion requirement and number of the resolvable paths, we can obtain a performance limit under the optimal condition as a reference. 
Finally, the experimental results of outage probability are obtained by Monte-Carlo method, and compared with the theoretical results calculated by the closed-from expression of the lower bound.
\end{abstract}

\section{Introduction}
Research towards developing \gls{6G} wireless communication networks is currently an emerging hot research topic globally, due to its huge potential in supporting various applications, such as  extended reality \cite{Xiao2023XR} and massive ultra-low latency communications \cite{Zhang2024Multiple}. 
Compared to \gls{5G} performance metrics, one of the challenges that \gls{6G} will face to is to satisfy the diverse and high-quality communication requirements in high user-terminal mobility scenarios, such as high-speed railways, unmanned aerial vehicles, and low earth orbit satellites \cite{Xu2023}. 
In fact, high-mobility broad band signal transmission over wireless channels imposes challenges that need to overcome frequency selectivity and time selectivity in the presence of fading variations. 
They are caused by excessive bandwidth and Doppler spread, compared to the channel coherence bandwidth and time, respectively. 

\Gls{OFDM}, which is widely used in \gls{5G} and its unprecedented systems, is a robust technique against fading frequency selectivity,  however its transmission performance is severely degraded due to the Doppler shift that destroys the orthogonality between subcarriers. 
In order to eliminate the detrimental effects caused by the doubly dispersive channels, a new modulation technique referred to as \gls{OTFS} modulation, with which the equivalent channel is defined in the \gls{DD} domain, is proposed in \cite{Hadani2017}. 
It has been widely recognized as a promising approach to reliable and robust communications in high user-terminal mobility scenarios\cite{Wei2021a}. 
Different from \gls{OFDM}, \gls{OTFS} processes the information symbols in the \gls{DD} domain, thus transforms the channels suffering from the double selectivity in the \gls{TF} domain into  quasistatic channels in the \gls{DD} domain exhibiting sparse and hence separable properties. 
Therefore, \gls{OTFS} requires only a lower channel estimation overhead than \gls{OFDM}. 
It is worth noting that modulating information symbols are defined in the DD domain, instead of TF domain, and hence each symbol in an OTFS frame experiences the effect of the full fluctuations in the \gls{TF} channel. 
Therefore, OTFS has the potential to exploit the full channel diversity, yielding better error performance in the high user-terminal mobility scenarios.

It is worth mentioning the representative researches aiming at realizing the practical \gls{OTFS} system.
\Gls{DD} channel response needs to be estimated for \gls{OTFS} signal detection at the receiver. 
In \cite{Raviteja2019a}, pilot-aided channel estimation techniques are studied. 
Furthermore, detection algorithms have also been proposed for \gls{OTFS} signal detection.  
Raviteja \emph{et al.} \cite{Raviteja2018a} proposes a low-complexity detector based on \gls{MP} algorithm. 
They estimate inter-symbol interference component as a Gaussian variable to reduce the detection complexity. 

Notice that current OTFS researches almost focus on lossless communications. 
However, a big tendency of wireless communication concept creation is a joint design with \gls{AI} \cite{Lin2024SF, Zhao2023Traffic, Fu2024Scalable, Lin2024SIC}, where the recovered information should not necessarily be lossless. 
As long as the final decision made by AI is correct, lossy recovery at the decoder output  still can be effectively exploited \cite{Lin2021Cooperative,Tad2024Two}. 
Therefore, lossy communications have a great potential for AI-oriented communications in \gls{6G}. 
Even though the previous contributions mentioned above have provided stringent framework for practical \gls{OTFS} system designs, the outage probability of \gls{OTFS} has not been thoroughly analyzed, especially for lossy communications. 

Due to the random channel variation in the \gls{DD} domain, the instantaneous channel capacity may not always support lossless transmission of the information sequence, in which case the information recovered at the receiver may well be lossy. 
Notice that it is challenging to derive the explicit expression of distortion under the instantaneous channel capacity constraints. We can utilize equivalent source coding based on Shannon’s lossy source-channel separation theorem to equivalently determine the distortion corresponding to the channel capacity \cite{Shannon1959}. Hence, determining the outage performance of \gls{OTFS} in lossy communications is of significant importance.

Given the background described above, we are motivated to investigate the outage probability of \gls{OTFS} in lossy communications. The main contributions of this paper include: 
1) Employing an equivalent source coding method based on Shannon’s lossy source-channel separation theorem for determining the \gls{SNR} corresponding to the distortion requirement with the aim of calculating the outage probability; 
2) Deriving the exact expression for the outage probability with \gls{OTFS} in lossy communications, given the distortion on requirement and resolvable path number; 
3) Deriving the lower bound of the outage probability without requiring heavy computational complexity.

\section{System Model}\label{sec:Principle}
This section provides a general description of the \gls{DD} domain channel model and the vector form representation of the \gls{OTFS} channel in the \gls{DD} domain, and then introduces the achievable capacity with the considered system model.

\subsection{Channel Model}
\label{section:A}
The received signal $r(t)$ transmitted from a transmitter with an input signal $s(t)$ over the \gls{DD} domain channel is given by
\begin{align}
r(t)=\iint h(\tau,\nu)s(t-\tau)e^{j2\pi\nu(t-\tau)}d\nu d\tau+w(t).
\end{align}
Assume that $w(t)$ is an \gls{AWGN} with one-sided \gls{PSD} $N_0$. Only a few parameters are often needed to model the channel in the \gls{DD} domain, since there are typically only a small number of reflectors in the channel with associated delays and Doppler shifts. Due to the sparsity of the channel representation, the response \emph{h(\text{$\tau$}, \text{$\nu$})} is expressed in the following form:
\begin{align}
h(\tau,\nu)=\sum_{i\operatorname{=}1}^Ph_i\delta(\tau-\tau_i)\delta(\nu-\nu_i),
\end{align}
where \emph{\text{$\delta(\cdot)$}} is the Dirac delta function, \emph{P} is the number of independent resolvable paths, and $h_i$, $\tau_i$ and $\nu_i$ denote, respectively, the complex fading coefficient, delay and Doppler shift corresponding to the \emph{i}-th path. 

We assume that a \gls{CP} of length $L_{\mathrm{CP}}$ is appended to the sequence to be sent before transmission. 
Specifically, we denote the delay and the Doppler indice by $l_i$ and $k_i$, where $\tau_i=\frac{l_i}{M\Delta f}$ and $\nu_i=\frac{k_i}{NT}$.
Specifically, with $0\leq l_i \leq l_{max}$, $l_{max}$ is usually not greater than $L_{\mathrm{CP}}$, i.e., the maximum delay of the channel is $\tau_{max}=l_{max}T/{M}$. For the same reason, $-k_{max}\leq k_i\leq k_{max}$ is assumed, i.e., the maximum Doppler shift is $\nu_{max}=\frac{k_{max}}{NT}$. The effect of the fractional delays in typical wide-band systems can be ignored because the sampling period $\frac{1}{M\Delta f}$ in the delay domain is in common small. 
Besides, the effects of fractional Doppler shifts from the nearest Doppler grid can be mitigated by adding \gls{TF} domain windows, when we assume neither receive filter in the Doppler domain nor use sampling frequency higher than the Nyquist sampling rate \cite{Raviteja2018}. 
Therefore, the results derived in this paper can be straightforwardly extended to the fractional delay and Doppler shifts case.

Let $\mathbf{h}=\left[h_1,h_2,\ldots,h_P\right]^\mathrm{T}$ denote the channel coefficient vector of size $P\times1$, assuming the elements in $\mathbf{h}$ are \gls{i.i.d.} complex Gaussian random variables. 
Since the channel coefficients have uniform power delay and Doppler profile, $h_i$ for $1\leq i \leq P$ has mean $\mu$ and variance $1/(2P)$ per dimension and does not depend on the delay or Doppler shifts. In particular, with $\mu=0$, $\vert h_i\vert$ follows the Rayleigh distribution.

\subsection{Vector Form Representation of OTFS}
Let $M$ represent number of subcarriers and $N$ number of slots per \gls{OTFS} frame\cite{Hadani2017}. $\mathbf{F}_N$ is the normalized discrete Fourier transform matrices of size $N\times N$. 
We then define $\mathbf{y}$ and $\mathbf{x}$ to represent the \gls{DD} domain transmitted and received symbol vectors, respectively. 
$\mathbf{w}$ is equivalent sample vector of \gls{AWGN} in \gls{DD} domain with one-sided \gls{PSD} of $N_0$. The input-output relationship of \gls{OTFS} in the \gls{DD} domain is \cite{Chong2022a}
\begin{align}
\mathbf{y}=\mathbf{H}_\mathrm{DD}\mathbf{x}+\mathbf{w},
\end{align}
where $\mathbf{H}_\mathrm{DD}$ is the effective \gls{DD} domain channel matrix
\begin{align}
\mathbf{H}_{\mathrm{DD}}=\sum_{i=1}^{P}h_{i}\left(\mathbf{F}_{N}\otimes\mathbf{I}_{M}\right)\mathbf{\Pi}^{l_{i}}\mathbf{\Delta}^{k_{i}}\left(\mathbf{F}_{N}^{\mathrm{H}}\otimes\mathbf{I}_{M}\right),\label{con:HDD}
\end{align}
with $\mathbf{\Pi}$ being a forward cyclic shift permutation matrix describing the delay effect, and $\boldsymbol{\Delta}=\operatorname{diag}\{\alpha^0,\alpha^1,\ldots,\alpha^{MN-1}\}$ being a diagonal matrix describing the Doppler effect with $\alpha=e^{\frac{j2\pi}{MN}}$\cite{Chong2022a}.

We now focus on the analysis of the achievable capacity, assuming a \gls{2D} Gaussian codebook with an average symbol energy $E_s$ for transmission. On the basis of the vector form representation of the input-output relationship in the \gls{DD} domain, the normalized achievable capacity is calculated as \cite{Chong2022a}
\begin{small}
\begin{align}
C=I(\mathbf{y};\mathbf{x})=\frac1{MN}\mathrm{log}_2\det\left(\mathbf{I}_{MN}+\frac{E_s}{N_0}\mathbf{H}_{\mathrm{DD}}^\mathrm{H}\mathbf{H}_{\mathrm{DD}}\right).\label{con:Iyx}
\end{align}
\end{small}Based on the previous analysis of the system model, outage performance analysis is presented in the next section.
\section{Outage Performance Analysis}\label{sec:Coding}
This section gives a definition for the outage probability of OTFS in lossy communications, and then derives the lower bound of the outage probability.
\subsection{Rate-Distortion Analysis}
Assume $T\times \Delta f=1$ throughout this paper, i.e., \gls{OTFS} is critically sampled for all pulse shaping waveforms. For an \gls{OTFS} system with $M$ subcarriers and $N$ time slots, the length of original binary sequence required to be modulated is $L=M\times N\times K$ per frame, with $2^K$ being the modulation order. 
Because the channel variation due to fading is random in time, the instantaneous channel capacity may not be able to support lossless transmission of the original sequence of $L$ bits. 
In this case, the information recovered at the receiver is lossy. To equivalently evaluate the distortion of the recovered information sequence, we adopt Shannon's lossy source-channel separation theorem for performance analysis. Based on Shannon's lossy source-channel separation theorem, the original sequence of $L$ bits is first lossy-compressed  into a codeword of length $S$, $S=M\times N\times R$,  where $R$ is the target rate. 
Therefore, the lossy compression rate can be represented as $R_S=\frac{MNR}{MNK}=\frac{R}{K}$.

In each OTFS frame, there are $M$ subcarriers and $N$ time slots, and hence the corresponding block in the DD domain can be divided into $M \times N$ resource sub-blocks. 
Each DD resource sub-block can allocate one symbol to carry information.
Assuming that the channel coefficient vector $\mathbf{h}$ is fixed within one OTFS frame, then the capacity of each DD resource sub-block averaged within one frame can be represented by $C$ as depicted in Fig. 1. 
Hence, the information that can be transmitted in one OTFS frame is $M \times N \times C$.
 Thus lossless transmission of the codeword of length $S$ is guaranteed whenever $S\leq M\times N\times C$, i.e., $R\leq C$. Then the receiver can recover a lossy sequence of length $L$ with distortion $D$ according to the codeword\footnote{To achieve the performance of Shannon's lossy source-channel separation theorem, the code length has to be long enough. 
Therefore, the achievable rate-distortion region derived in this paper provides a lower bound.}.
 
\begin{figure}[!t]
\centering \includegraphics[width=1\figwidth]{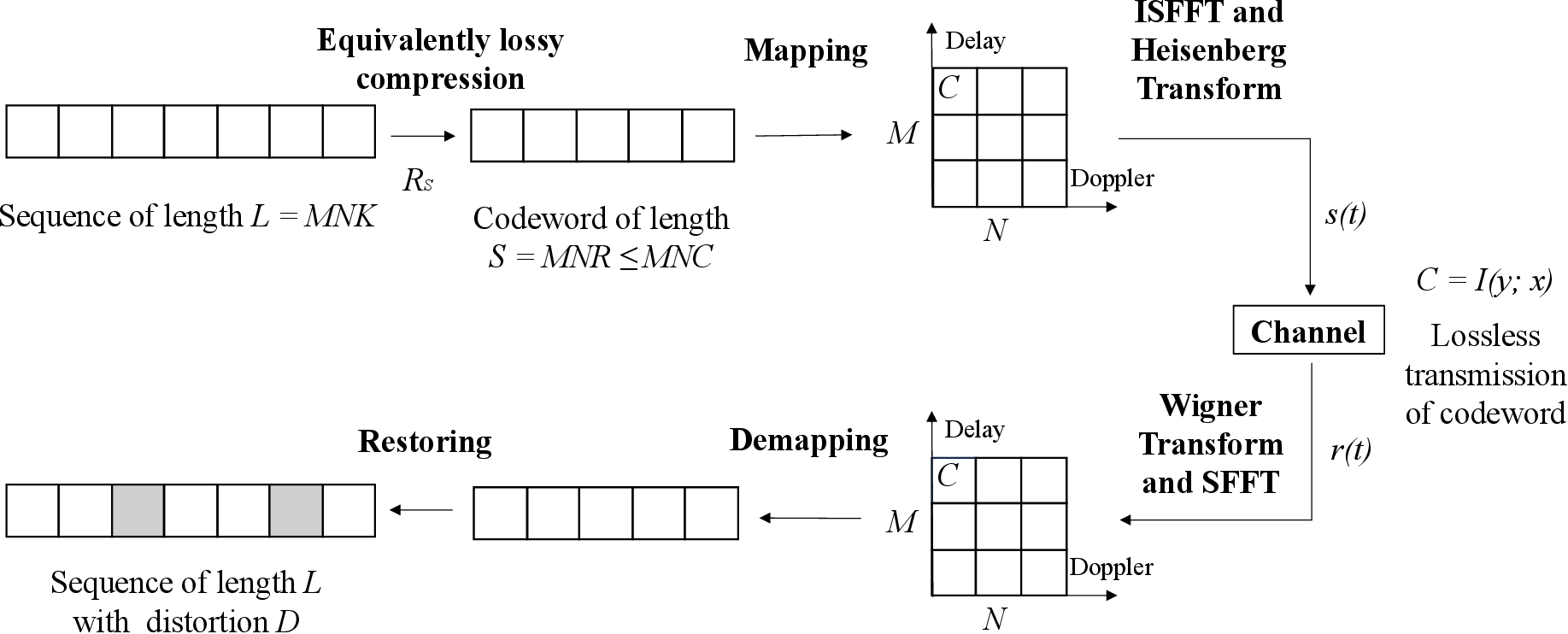}
\caption{Equivalent system model of lossy transmissions with OTFS based on lossy source-channel separation theorem.}
\label{fig:a_RD}
\vspace{-2em}
\end{figure}

The equivalent system model of \gls{OTFS} in lossy communications presented in \fref{fig:a_RD}. After \gls{ISFFT} and Heisenberg transform\cite{Hadani2017}, the codeword is transmitted over the channel. Assuming that the signal $s(t)$ can be transmitted losslessly in the channel. At the receiver, the received signal $r(t)$ is transformed to the \gls{DD} domain through the Wigner transform and \gls{SFFT}, prior to symbol demodulation. Then, the sequence is restored to its original length $L$ with distortion $D$. According to the rate-distortion theorem, the distortion can be calculated as
\begin{align}
D=H_b^{-1}\left(1-R/K\right),\label{con:RateDis}
\end{align}
where $H_b{(\cdot)}$ is the binary entropy function. Furthermore, for a binary source, the Hamming distortion is equivalent to the \gls{BER}, i.e.,
$\gls{BER}=D$.\par
Based on (\ref{con:Iyx}) and (\ref{con:RateDis}), the outage probability is defined as
\begin{small}
\begin{align}
P_{\mathrm{out}}&=\mathrm{Pr}\left\{\frac{1}{MN}\mathrm{log}_{2}\operatorname*{det}\left(\mathbf{I}_{MN}+\frac{E_{s}}{N_{0}}\mathbf{H}_{\mathrm{DD}}^{\mathrm{H}}\mathbf{H}_{\mathrm{DD}}\right)<R\right\}\nonumber\\
&=\mathrm{Pr}\left\{H_{b}^{-1}\!\!\left[1\!-\!\frac{1}{M\!N\!K}\mathrm{log}_{2}\!\det\!\left(\!\mathbf{I}_{M\!N}\!+\!\frac{E_{s}}{N_{0}}\mathbf{H}_{\mathrm{DD}}^{\mathrm{H}}\mathbf{H}_{\mathrm{DD}}\!\right)\!\right]\!>\!D\right\}\!.\label{con:poutdef}
\end{align}
\end{small}It is obvious that the outage probability of the \gls{OTFS}  in lossy communications is closely related to the effective \gls{DD} domain channel matrix $\mathbf{H}_{\mathrm{DD}}$. In the next subsection we will investigate the properties of $\mathbf{H}_{\mathrm{DD}}^{\mathrm{H}}\mathbf{H}_{\mathrm{DD}}$ and derive a lower bound of the outage probability correspondingly.
\subsection{Lower Bound Derivation}
Note that the matrix $\mathbf{H}_{\mathrm{DD}}^{\mathrm{H}}\mathbf{H}_{\mathrm{DD}}$ is a decomposable Hermitian matrix, i.e., it has nonnegative eigenvalues\cite{Chong2022a}. Based on (\ref{con:HDD}), $\mathbf{H}_{\mathrm{DD}}^{\mathrm{H}}\mathbf{H}_{\mathrm{DD}}$ can be written as
\begin{align}
\mathbf{H}_{\mathrm{DD}}^{\mathrm{H}}\mathbf{H}_{\mathrm{DD}}&=\left[\sum_{i=1}^Ph_i^*(\mathbf{F}_N\otimes\mathbf{I}_M)(\mathbf{\Delta}^{k_i})^\mathrm{H}(\mathbf{\Pi}^{l_i})^\mathrm{H}\big(\mathbf{F}_N^\mathrm{H}\otimes\mathbf{I}_M\big)\right]\nonumber\\
&\quad\times\left[\sum_{i=1}^Ph_i(\mathbf{F}_N\otimes\mathbf{I}_M)\mathbf{\Pi}^{l_i}\mathbf{\Delta}^{k_i}\big(\mathbf{F}_N^\mathrm{H}\otimes\mathbf{I}_M\big)\right]\nonumber\\
&=\sum_{i=1}^{P}|h_{i}|^{2}\mathbf{I}_{MN}+\sum_{i=1}^{P}\sum_{\substack{i'=1\\i'\neq i}}^{P}h_{i}^{*}h_{i'}(\mathbf{F}_{N}\otimes\mathbf{I}_{M})\mathbf{\Delta}^{-k_{i}}\nonumber\\
&\quad\mathbf{\Pi}^{l_{i'}-l_{i}}\mathbf{\Delta}^{k_{i'}}(\mathbf{F}_{N}^{\mathrm{H}}\otimes\mathbf{I}_{M}).
\end{align}
To simplify the expression, we let $\mathbf{H}_{\mathrm{A}}=\sum_{i=1}^{P}|h_{i}|^{2}\mathbf{I}_{MN}$ and 
\begin{small}
\begin{align}
\mathbf{H}_{\mathrm{B1}}&= \sum_{i=1}^{P-1} \sum_{\substack{i'=i+1,\\
l_{i'} = l_i}}^P \left(\mathbf{F}_N \otimes \mathbf{I}_M\right) \left( h_i^* h_{i'}\boldsymbol{\Lambda} + h_i h_{i'}^* \boldsymbol{\Lambda}^{\mathrm{H}}\right) \left(\mathbf{F}_N^{\mathrm{H}} \otimes \mathbf{I}_M\right),\nonumber
\\
\mathbf{H}_{\mathrm{B2}}&= \sum_{i=1}^{P-1} \sum_{\substack{i'=i+1,\\
l_{i'} \neq l_i}}^P \left(\mathbf{F}_N \otimes \mathbf{I}_M\right) \left( h_i^* h_{i'}\boldsymbol{\Lambda} + h_i h_{i'}^* \boldsymbol{\Lambda}^{\mathrm{H}}\right) \left(\mathbf{F}_N^{\mathrm{H}} \otimes \mathbf{I}_M\right),
\end{align}
\end{small}where, $\boldsymbol{\Lambda}=\boldsymbol{\Delta}^{-k_{i}}\boldsymbol{\Pi}^{l_{i^{\prime}}-l_{i}}\boldsymbol{\Delta}^{k_{i^{\prime}}}$, hence $\mathbf{H}_{\mathrm{DD}}^{\mathrm{H}}\mathbf{H}_{\mathrm{DD}}=\mathbf{H}_{\mathrm{A}}+\mathbf{H}_{\mathrm{B1}}+\mathbf{H}_{\mathrm{B2}}$. Then, we can obtain the theoretical outage probability of \gls{OTFS} in lossy communications, as
\begin{align}
P_{out}=\iiint_Qdh_{\mathrm{Re},1}dh_{\mathrm{Im},1}\cdotp\cdotp\cdotp dh_{\mathrm{Re},i}dh_{\mathrm{Im},i},\label{con:poutA}
\end{align}
where, $Q$ is the region enclosed by 
\begin{small}
\begin{align}
H_b^{-1}\!\left\{\!1\!-\!\frac1{MNK}\log_2\det\!\left[\mathbf{I}_{MN}\!+\frac{E_S}{N_0}(\mathbf{H}_{\mathrm{A}}\!+\!\mathbf{H}_{\mathrm{B1}}\!+\!\mathbf{H}_{\mathrm{B2}})\right]\right\}\!>\!D.
\end{align}
\end{small}

Notice that it is difficult to obtain a concise closed-form expression of (\ref{con:poutA}). Although $P_{out}$ can be calculated using numerical algorithms, its computational complexity is quite heavy. 
However, it is found that the calculation of $P_{out}$ can be simplified by deriving a lower bound, for which we propose two propositions.\par
\emph{Proposition 1}:
\begin{small}
\begin{align}
\mathrm{det}\!\left(\mathbf{I}_{M N}\!+\!\frac{E_s}{N_0}(\mathbf{H}_{\mathrm{A}}\!+\!\mathbf{H}_{\mathrm{B1}}) \right)\!
\le\! \mathrm{det}\!\left(\!\mathbf{I}_{M N}\!+\!\frac{E_s}{N_0} \mathbf{H}_{\mathrm{A}}\!\right)\!.
\end{align}
\end{small}The proof of \emph{Proposition 1} is presented in Appendix A.\par
\emph{Proposition 2}:
\begin{small}
\begin{align}
\mathrm{det}\left(\mathbf{I}_{M N}+\frac{E_s}{N_0} (\mathbf{H}_{\mathrm{A}}+\mathbf{H}_{\mathrm{B1}} +\mathbf{H}_{\mathrm{B2}}) \right) 
\le \nonumber\\
\mathrm{det}\left(\mathbf{I}_{M N}+\frac{E_s}{N_0} (\mathbf{H}_{\mathrm{A}}+\mathbf{H}_{\mathrm{B1}}) \right).
\end{align}
\end{small}The proof of \emph{Proposition 2} is presented in Appendix B.\par
Combining \emph{Proposition 1} and \emph{Proposition 2}, we have
\begin{small}
\begin{align}
\mathrm{det}\left(\mathbf{I}_{M N}+\frac{E_s}{N_0} \mathbf{H}_{\mathrm{DD}}^{\mathrm{H}} \mathbf{H}_{\mathrm{DD}} \right)\! \le \mathrm{det}\left(\mathbf{I}_{M N}+\frac{E_s}{N_0} \mathbf{H}_{\mathrm{A}} \right).
\end{align}
\end{small}It matches our intuition that the capacity becomes maximum when the energy from all paths is fully combined without loss. This can give a performance limit under the optimal condition as a reference.
Moreover, based on (\ref{con:poutdef}), we can deduce the lower bound of $P_{out}$:
\begin{small}
\begin{align}
P_{\text {out }} & \geq \Pr \left\{H_b^{-1}\left[1-\frac{1}{MNK} \log_2 \mathrm{det}\!\left(\mathbf{I}_{M N}+\frac{E_s}{N_0} \mathbf{H}_{\mathrm{A}} \!\right)\right]\!>D\right\} \nonumber\\
&= \Pr \left\{H_b^{-1}\left[1-\frac{1}{K} \log _2\left(1+\frac{E_s}{N_0} \sum_{i=1}^P\left|h_i\right|^2\right)\right]>D\right\} \nonumber\\
&=\Pr\left\{\sum_{i=1}^P\left|h_i\right|^2<\frac{2^{K-K H_b(D)}-1}{E_s / N_0}\right\}.\label{con:prhi}
\end{align}
\end{small}\par
To simplify the derivation, we define vector $\boldsymbol{h^{\prime}}=[h^{\prime}_1,h^{\prime}_2,\ldots,h^{\prime}_P]^{\mathrm{T}}$ with $P$ elements, where $h^{\prime}_i=\sqrt{2P}h_i$. As discussed in Section \ref{section:A}, $h_i$ follows for any $i$ an i.i.d complex Gaussian distribution, and hence $h^{\prime}_i$ for $1\leq i \leq P$ are \gls{i.i.d.} complex Gaussian random variables with zero mean and unit variance per dimension. 
Therefore, $\sum_{i=1}^P|h^{\prime}{}_i|^2$ follows a Chi-squared distribution with $2P$ degrees of freedom. In this way, (\ref{con:prhi}) can be further written as 
\begin{small}
\begin{align}
P_{out}\geq\!\Pr\left\{\sum_{i=1}^P|h^{\prime}{}_i|^2<\frac{2P(2^{K-KH_b(D)}-1)}{E_s/N_0}\right\}.
\end{align}
\end{small}Eventually, from the \gls{PDF} of the Chi-square distribution\cite[Section 5.8]{Ross2000}, we can calculate
\begin{small}
\begin{align}
P_{\mathrm{out}}\!\geq\!1\!-\!\exp\!\left(\!-\frac{P(2^{K-KH_b(D)}-1)}{E_s/N_0}\!\right)\!\sum_{i=0}^{P-1}\frac{P^i(2^{K-KH_b(D)}-1)^i}{i!(E_s/N_0)^i}.\label{con:poutresult}
\end{align}
\end{small}

\section{Performance Evaluation}\label{sec:Evaluation}
In this section, we employed the Monte-Carlo method\cite[Section 2.3]{Barbu2020} to obtain the outage probability experimentally where randomly generated $h_i$ is used to evaluate the theoretical outage expressed by the multi-fold integral given by (\ref{con:poutA}). They were then compared to the lower-bound outage probability derived in the previous session. Without loss of generality, we set ${\Delta f}=15\mathrm{KHz}$ and $f_c=4\mathrm{GHz}$, and use \gls{MRC} at the detector. 
The channel coefficients are randomly generated based on the uniform power delay and Doppler profile, as discussed in Section \ref{section:A}. To facilitate comparison of the results, we fixed $l_i$ and $k_i$ in the  ranges $[0,8]$ and $[-8,8]$, respectively. Therefore, the effect of moving speed on the experimental results can be neglected.

\begin{figure}[!t]
\centering \includegraphics[width=1\figwidth]{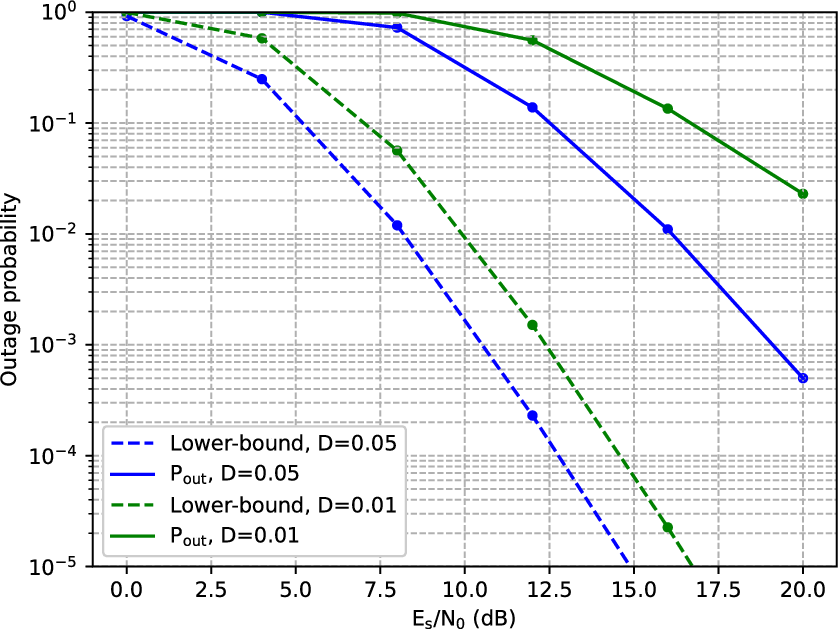}
\vspace{-1.5em}
\caption{The outage probability with $M=N=16$, $P=5$.}
\label{fig:P5M16}
\vspace{-1em}
\end{figure}

We first demonstrate the lower-bound outage probability and the experimental outage probability for different frame lengths. \fref{fig:P5M16} illustrates the curves of the lower bound and the experimental of the outage probability versus $E_s/N_0$ for $M=N=16$ and $P=5$. 
It is clearly seen that the outage probability decreases with greater distortion $D$, with the same $E_s/N_0$.

\begin{figure}[!t]
\centering \includegraphics[width=1\figwidth]{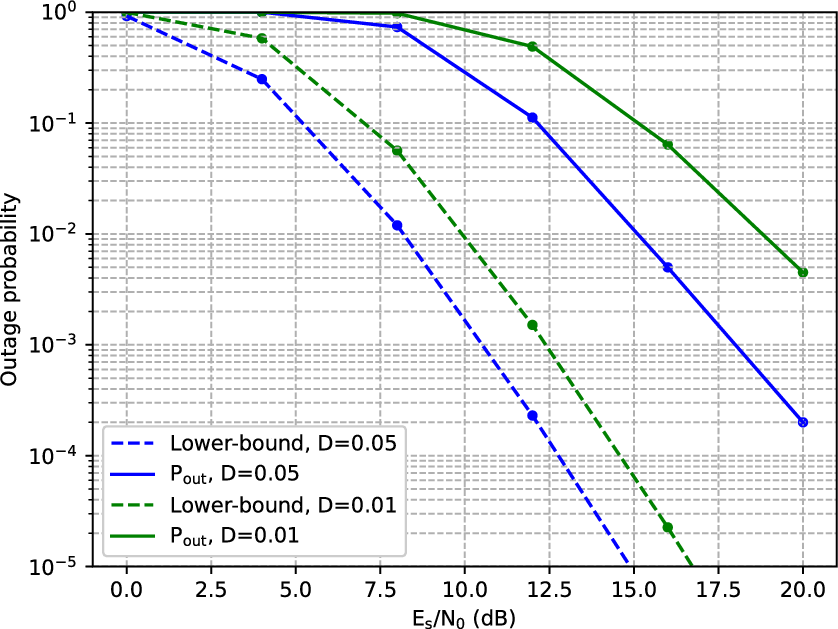}
\vspace{-1.5em}
\caption{The outage probability with $M=N=32$, $P=5$.}
\label{fig:P5M32}
\end{figure}

\fref{fig:P5M32} shows the lower bound and the experimental outage probability with $M=N=32$. 
In high $E_s/N_0$ value range, the experimental outage probability curve in \fref{fig:P5M32} has a significantly steeper slope than that in \fref{fig:P5M16}, and is also closer to the lower bound curve. 
This is because the rate-distortion theorem requires the code length needs to be infinite, so the longer the frame length the smaller the gap. 
Thus, the longer the frame length, the closer the outage probability is to the actual value. 
It is also worth noting that in the high $E_s/N_0$ region, the slope of the experimental outage probability and the corresponding lower bound are essentially the same.

\begin{figure}[!t]
\centering \includegraphics[width=1\figwidth]{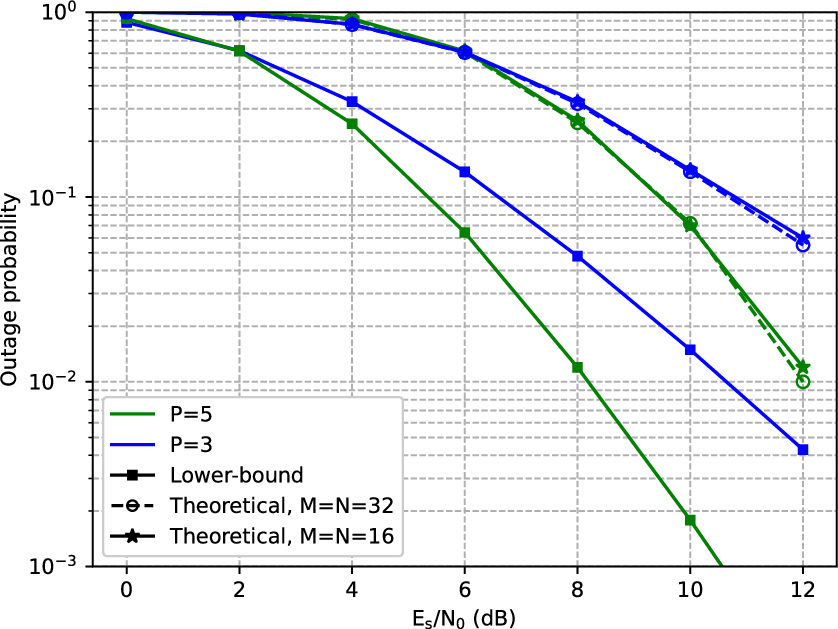}
\vspace{-1.5em}
\caption{The outage probability of \gls{OTFS} in lossy communications with different numbers of paths and frame lengths, where the distortion $D=0.05$.}
\label{fig:compare}
\vspace{-1em}
\end{figure}

\fref{fig:compare} illustrates the theoretical curves of the \gls{OTFS} outage probability for different resolvable paths and frame lengths. 
It is clear that as number of resolvable paths increases, the outage probability decreases under the same conditions. In addition, the outage probability for various frame lengths coincides each other over a relatively large value range of $E_s/N_0$. 
However, the outage probability at $M=N=32$ is slightly lower than that at $M=N=16$. 
The difference of the outage probability becomes larger as $E_s/N_0$ increases.

\section{Conclusion}\label{sec:Conclusion}

We have analyzed the outage probability of \gls{OTFS} in lossy communications. We start from the rate-distortion analysis of \gls{OTFS} in lossy communications based on Shannon’s lossy source-channel separation theorem. 
Then, we derived an exact expression for the outage probability, expressed by a multi-fold integral with respect to the PDFs of the instantaneous channel coefficients, and hence numerical calculation for the multi-fold integral requires heavy computational complexity. 
To simplify the calculations, we obtained a lower bound on the outage probability. 
It has been found that the numerical results are consistent in their curve tendency to the lower bound analysis, and the longer the frame length, the closer the actual outage probability obtained by the Monte-Carlo integral.
\appendices
\section{Proof of \emph{Proposition 1}}
Consider
\begin{small}
\begin{align}
\mathbf{H}_{\mathrm{B1}}&= \sum_{i=1}^{P-1} \sum_{\substack{i'=i+1,\\
l_{i'} = l_i}}^P \left(\mathbf{F}_N \otimes \mathbf{I}_M\right) \left( h_i^* h_{i'}\boldsymbol{\Lambda} + h_i h_{i'}^* \boldsymbol{\Lambda}\right) \left(\mathbf{F}_N^{\mathrm{H}} \otimes \mathbf{I}_M\right)\nonumber\\
  &= \sum_{i=1}^{P-1} \sum_{\substack{i'=i+1,\\
l_{i'} = l_i}}^P \left(\mathbf{F}_N \otimes \mathbf{I}_M\right) \left( h_i^* h_{i'} \boldsymbol{\Delta}^{k_{i'}-k_i} + h_i h_{i'}^* \boldsymbol{\Delta}^{k_i-k_{i'}} \right) \nonumber\\&\quad\left(\mathbf{F}_N^{\mathrm{H}} \otimes \mathbf{I}_M\right),\label{eq:H_B1}
\end{align}
\end{small}where, $\boldsymbol{\Lambda}=\boldsymbol{\Delta}^{-k_{i}}\boldsymbol{\Pi}^{l_{i^{\prime}}-l_{i}}\boldsymbol{\Delta}^{k_{i^{\prime}}}$, when $l_{i'} = l_i$, $\boldsymbol{\Lambda}=\boldsymbol{\Delta}^{k_i-k_{i'}}$. 
For the ease of derivation, let $k_{i',i}=k_{i'}-k_i \neq 0 $ and $h_i^* h_{i'}=|h_{i,i'}|e^{j\theta_{i,i'}}$. 
Then, we further analyze $\boldsymbol{\Theta}=h_i^* h_{i'} \boldsymbol{\Delta}^{k_{i'}-k_i} + h_i h_{i'}^* \boldsymbol{\Delta}^{k_i-k_{i'}} $, as
\begin{align}
\boldsymbol{\Theta}
&= |h_{i,i'}|e^{j\theta_{i,i'}} \mathrm{diag}\left\{e^{j\frac{2\pi \times 0}{MN}k_{i',i}}, \cdots, e^{j\frac{2\pi (MN-1)}{MN}k_{i',i}}\right\} \nonumber\\
&\quad + |h_{i,i'}|e^{-j\theta_{i,i'}} \mathrm{diag}\left\{e^{-j\frac{2\pi \times 0}{MN}k_{i',i}}, \cdots, e^{-j\frac{2\pi (MN-1)}{MN}k_{i',i}}\right\} \nonumber\\
&=  |h_{i,i'}|\mathrm{diag}\left\{e^{j\left(\frac{2\pi \times 0}{MN}k_{i',i}+\theta_{i,i'}\right)}+e^{-j\left(\frac{2\pi \times 0}{MN}k_{i',i}+\theta_{i,i'}\right)}, \cdots, \right. \nonumber\\
&\qquad \qquad \left. e^{j\left(\frac{2\pi (MN-1)}{MN}k_{i',i}+\theta_{i,i'}\right)}+e^{-j\left(\frac{2\pi (MN-1)}{MN}k_{i',i}+\theta_{i,i'}\right)}\right\} \nonumber\\
&=  2|h_{i,i'}|\mathrm{diag}\left\{\cos{\left(\frac{2\pi \times 0}{MN}k_{i',i}+\theta_{i,i'}\right)}, \cdots, \right. \nonumber\\
&\qquad \qquad \left. \cos{\left(\frac{2\pi (MN-1)}{MN}k_{i',i}+\theta_{i,i'}\right)}\right\}\nonumber\\
&=  2|h_{i,i'}|\mathrm{diag}\left\{\beta_0, \cdots, \beta_{MN-1}\right\},  \label{eq:H_B1_diag}
\end{align}
where, $\beta_b=\cos{\left(\frac{2\pi b}{MN}k_{i',i}+\theta_{i,i'}\right)}$.\par
For the convenience of the \gls{FFT}, $M$ and $N$ are integer powers of 2 in most systems. Assume $k_{i',i}=a_1 \times a_2$, where, $a_1$ is odd, $a_2=2^q, q\in\{0,1,\cdots,\log_2{N}-1\}$. Since $k_{i',i}<N$, $N/a_2$ is integer power of 2 as well. We have
\begin{align}
\beta_{b+\frac{MN}{2a_2}}&=\cos{\left(\frac{2\pi \left(b+\frac{MN}{2a_2}\right)}{MN}k_{i',i}+\theta_{i,i'}\right)}\nonumber\\
&=\cos{\left(\frac{2\pi b}{MN}k_{i',i}+\pi a_1 +\theta_{i,i'}\right)}\nonumber\\
&=-\cos{\left(\frac{2\pi b}{MN}k_{i',i}+\theta_{i,i'}\right)}=-\beta_b.
\end{align}
Therefore, $\mathrm{diag}\left\{\beta_0, \cdots, \beta_{MN-1}\right\}$ can be divided into $2a_2$ groups, and the odd group and the next even group have the same value and opposite signs, as
\begin{small}
\begin{align}
\quad \mathrm{diag}\big\{\underbrace{\beta_0, \cdots,\beta_{\frac{MN}{2a_2}-1}}_\text{1st group},
\underbrace{\beta_{\frac{MN}{2a_2}},\cdots, \beta_{\frac{2MN}{2a_2}-1}}_\text{2nd group}, \cdots, \nonumber\\
\underbrace{\beta_{\frac{(2a_2-1)MN}{2a_2}},\cdots,\beta_{MN-1}}_{2a_2\text{-th group}}   \big\} \nonumber\\
=\mathrm{diag}\big\{\underbrace{\beta_0, \cdots,\beta_{\frac{MN}{2a_2}-1}}_\text{1st group},
\underbrace{-\beta_0, \cdots,-\beta_{\frac{MN}{2a_2}-1}}_\text{2nd group}, \cdots \big\},
\end{align}
\end{small}i.e., half of the numbers in $\mathrm{diag}\left\{\beta_0, \cdots, \beta_{MN-1}\right\}$ are opposite the other half.\par
Combining \eqref{eq:H_B1} and \eqref{eq:H_B1_diag}, we have
\begin{small}
\begin{align}
\mathbf{I}_{M N}+\frac{E_s}{N_0} (\mathbf{H}_{\mathrm{A}}+\mathbf{H}_{\mathrm{B1}}) =\left(\mathbf{F}_N \otimes \mathbf{I}_M\right) \boldsymbol{\Xi} \left(\mathbf{F}_N^{\mathrm{H}} \otimes \mathbf{I}_M\right),
\end{align}
\end{small}where
\begin{align}    
\boldsymbol{\Xi}&=\xi_1  \mathbf{I}_{M N}
+ \mathrm{diag}\left\{\xi_{2,0}, \cdots, \xi_{2,MN-1}\right\}, \\
\xi_1&=1+\frac{E_s}{N_0}\sum_{i=1}^P\left|h_i\right|^2,\\
\xi_{2,b}&=\sum_{i=1}^{P-1} \sum_{\substack{i'=i+1,\\
l_{i'} = l_i}}^P \frac{2E_s}{N_0}|h_{i,i'}| \beta_b .
\end{align}
Since $\boldsymbol{\Xi}$ is a diagonal matrix,
\begin{align}    
\mathrm{det}\left(\mathbf{I}_{M N}+\frac{E_s}{N_0} (\mathbf{H}_{\mathrm{A}}+\mathbf{H}_{\mathrm{B1}}) \right) &= \mathrm{det} (\boldsymbol{\Xi}) \label{eq:det_xii}\\
&=\prod_{b=0}^{MN-1} \left( \xi_1 + \xi_{2,b} \right). \label{eq:det_Xi}
\end{align}\par
Because $\beta_{b+\frac{MN}{2a_2}}=-\beta_b$, the product of the $b$-th and the $\left(b+\frac{MN}{2a_2} \right)$-th terms in \eqref{eq:det_Xi} becomes
\begin{align}
\left( \xi_1 + \xi_{2,b} \right) \left( \xi_1 + \xi_{2,b+\frac{MN}{2a_2}} \right)&= \left( \xi_1 + \xi_{2,b} \right) \left( \xi_1 - \xi_{2,b} \right)  \nonumber\\
&=\xi_1^2 - \xi_{2,b}^2 \le \xi_1^2,
\end{align}
Therefore, 
\begin{small}
\begin{align}
    \mathrm{det}\left(\mathbf{I}_{M N}+\frac{E_s}{N_0} (\mathbf{H}_{\mathrm{A}}+\mathbf{H}_{\mathrm{B1}}) \right) \le \xi_1^{MN} 
    \!=\! \mathrm{det}\left(\mathbf{I}_{M N}\!+\!\frac{E_s}{N_0} \mathbf{H}_{\mathrm{A}} \right).
\end{align}
\end{small}This finishes the proof of \emph{Proposition 1}.
\section{Proof of \emph{Proposition 2}}
Consider
\begin{align}
&\quad \mathrm{det}\left(\mathbf{I}_{M N}+\frac{E_s}{N_0} (\mathbf{H}_{\mathrm{A}}+\mathbf{H}_{\mathrm{B1}} +\mathbf{H}_{\mathrm{B2}}) \right) \nonumber\\
&=\mathrm{det}\left[ \left(\mathbf{F}_N \!\otimes\! \mathbf{I}_M\right) \boldsymbol{\Xi} \left(\mathbf{F}_N^{\mathrm{H}}\! \otimes \!\mathbf{I}_M\right)\!+\!\left(\mathbf{F}_N \!\otimes \!\mathbf{I}_M\right)\boldsymbol{\Omega} \left(\mathbf{F}_N^{\mathrm{H}} \!\otimes\! \mathbf{I}_M\right) \right]\nonumber\\
&=\mathrm{det}\left[ \left(\mathbf{F}_N \otimes \mathbf{I}_M\right) \left(\boldsymbol{\Xi} +\boldsymbol{\Omega} \right) \left(\mathbf{F}_N^{\mathrm{H}} \otimes \mathbf{I}_M\right)\right]\nonumber\\
&=\mathrm{det} \left(\boldsymbol{\Xi} +\boldsymbol{\Omega} \right) ,\label{eq:det_xioumu}
\end{align}
where
\begin{align}
\boldsymbol{\Omega}=\frac{E_s}{N_0} \sum_{i=1}^{P-1} \sum_{\substack{i'=i+1,\\
l_{i'} \neq l_i}}^P \left( h_i^* h_{i'}\boldsymbol{\Lambda} + h_i h_{i'}^* \boldsymbol{\Lambda}^{\mathrm{H}}  \right).
\end{align}
Since $l_{i^{\prime}}\neq l_i<M$, $\Pi^{l_{i^{\prime}}-l_i}$ is a permutation matrix that will not be a identity array, i.e., all elements on its diagonal are zero. Moreover, $\boldsymbol{\Delta}^{-k_i}$ and $\boldsymbol{\Delta}^{k_{i^{\prime}}}$ are diagonal matrices, so all diagonal elements of $\boldsymbol{\Lambda}$ are zero, and therefore all diagonal elements of $\boldsymbol{\Omega}$ are zero.\par
Since $\mathbf{H}_{\mathrm{DD}}^{\mathrm{H}}\mathbf{H}_{\mathrm{DD}}$ is a semi-positive definite matrix and $\mathbf{I}_{M N}$ is a positive definite matrix, $\left(\mathbf{I}_{M N}+\mathbf{H}_{\mathrm{DD}}^{\mathrm{H}} \mathbf{H}_{\mathrm{DD}}\right)$ is a positive definite matrix. Because $\left(\boldsymbol{\Xi} +\boldsymbol{\Omega} \right) $ is a similarity matrix to $\left(\mathbf{I}_{M N}+\mathbf{H}_{\mathrm{DD}}^{\mathrm{H}} \mathbf{H}_{\mathrm{DD}}\right)$, $\left(\boldsymbol{\Xi} +\boldsymbol{\Omega} \right) $ is also a positive definite matrix. Moreover, $\boldsymbol{\Xi}$ is a diagonal matrix and all diagonal elements of $\boldsymbol{\Omega}$ are zero. Because the determinant of a semi-positive definite matrix is not greater than the product of its diagonal elements\cite[Theorem 11]{Rozanski2017}, we have
\begin{align}
\mathrm{det} \left(\boldsymbol{\Xi} +\boldsymbol{\Omega} \right) \le \mathrm{det} \left(\boldsymbol{\Xi}  \right) .\label{eq:det_le}
\end{align}
By combining \eqref{eq:det_xii}, \eqref{eq:det_xioumu} and \eqref{eq:det_le}, we have finished the proof of \emph{Proposition 2}.
\bibliographystyle{IEEEtran}
\bibliography{myreference}

\end{document}